\documentclass[12pt]{report}
\usepackage{latexsym,amsmath,amssymb,color}
\usepackage[dvips,colorlinks=true,linkcolor=blue]{hyperref}

\def\F{I\kern-.30em{F}}
\def\P{I\kern-.30em{P}}
\def\E{I\kern-.30em{E}}
\def\build#1_#2^#3{\mathrel{\mathop{\kern
0pt#1}\limits_{#2}^{#3}}}
\def\Sum{\displaystyle\sum}

\newcommand{\R}{\mathbb{R}}

\newcommand{\Z}{\mathbb{Z}}

\newcommand{\N}{\mathbb{N}}

\newcommand{\C}{\mathbb{C}}

\newcommand{\D}{\displaystyle}

\newcommand{\pro}{\mathbb{P}}

\newtheorem{theorem}{Theorem}[section]
\newtheorem{lemma}{Lemma}[section]
\newtheorem{proposition}{Proposition}[section]
\newtheorem{corollary}{Corollary}[section]

\newcommand{\Schr}{Schr{\"o}dinger}

\newcommand{\beq}{\begin{equation}}
\newcommand{\eeq}{\end{equation}}
\newcommand{\ba}{\begin{array}}
\newcommand{\ea}{\end{array}}
\newcommand{\bea}{\begin{eqnarray}}
\newcommand{\eea}{\end{eqnarray}}

\begin{document}

\begin{titlepage}

  \begin{center}

    {\bf AN OPTIMAL WEGNER ESTIMATE \\
     AND ITS APPLICATION TO THE GLOBAL CONTINUITY \\
     OF THE INTEGRATED DENSITY OF STATES \\
      FOR RANDOM SCHR{\"O}DINGER OPERATORS}

    \vspace{0.1 cm}

    \setcounter{footnote}{0}
    \renewcommand{\thefootnote}{\arabic{footnote}}

    {\bf Jean-Michel Combes \footnote{Centre de Physique Th{\'e}orique,
        CNRS Marseille, France} }

    \vspace{0.1 cm}

    {D{\'e}partement de Math{\'e}matiques \\
      Universit{\'e} du Sud, Toulon-Var \\
      83130 La Garde, FRANCE}

    \vspace{0.1 cm}

    {\bf Peter D.\ Hislop \footnote{Supported in part by NSF grant
        DMS-0503784.}}

    \vspace{0.1 cm}

    {Department of Mathematics \\
      University of Kentucky \\
      Lexington, KY 40506--0027 USA}

    \vspace{0.1 cm}

    {\bf Fr{\'e}d{\'e}ric Klopp}

    \vspace{0.1 cm}

    {L.A.G.A, Institut Galil{\'e}e\\
      Universit{\'e} Paris-Nord \\
      F-93430 Villetaneuse, FRANCE\\
      et\\
      Institut Universitaire de France}

  \end{center}

  \vspace{0.1 cm}

\begin{center}
  {\bf Abstract}
\end{center}

\noindent
We prove that the integrated density of states (IDS) of random \Schr\
operators with Anderson-type potentials on $L^2 ( \R^d)$, for $d \geq
1 $, is locally H{\"o}lder continuous at all energies with the same
H{\"o}lder exponent $0 < \alpha \leq 1$ as the conditional probability
measure for the single-site random variable.  As a special case, we
prove that if the probability distribution is absolutely continuous
with respect to Lebesgue measure with a bounded density, then the IDS
is Lipschitz continuous at all energies.  The single-site potential $u
\in L_0^\infty ( \R^d)$ must be nonnegative and compactly-supported.
The unperturbed Hamiltonian must be periodic and satisfy a unique
continuation principle.  We also prove analogous continuity results
for the IDS of random Anderson-type perturbations of the Landau
Hamiltonian in two-dimensions. All of these results follow from a new
Wegner estimate for local random Hamiltonians with rather general
probability measures.

\today

\end{titlepage}





\renewcommand{\thechapter}{\arabic{chapter}}
\renewcommand{\thesection}{\thechapter}

\setcounter{chapter}{1} \setcounter{equation}{0}

\section{Introduction and Main Results}\label{intro}

In this paper, we combine approaches of \cite{[CH1]} and \cite{[CHK1]}
to prove, as a special case, the Lipschitz continuity of the
integrated density of states (IDS) for random \Schr\ operators
$H_\omega = H_0 + V_\omega$, on $L^2 ( \R^d)$, for $d \geq 1$,
provided the conditional probability distribution for the random
variable at a single-site has a density in $L^\infty_0 (\R)$.  In
previous papers \cite{[CHK1],[CHKR]}, we proved global H{\"o}lder
continuity, for any order strictly less than one, of the IDS under the
same hypotheses on the single-site probability measure, and, in
\cite{[HKNSV]}, there was an improvement up to a logarithmic factor
(see below). It has long been expected that if the probability measure
of a single-site random variable has a bounded density with compact
support, then the IDS should be locally Lipschitz continuous at all
energies. This is known to be true if the single-site potential
satisfies a simple covering condition \cite{[CH1],[CH2]}. This result
is a special case of the continuity bound proved in this paper. We
prove that if the conditional probability measure is H{\"o}lder continuous
of order $0 < \alpha \leq 1$, then the IDS is H{\"o}lder continuous of
order $\alpha$ at all energies. Hence, the IDS has at least the same
continuity property as the conditional probability measure.  These
results follow from a Wegner estimate valid for a very general class
of probability measures. We refer to \cite{[CHK1]} for an introduction
to the problem and discussion of previous results.

The family of \Schr\ operators $H_\omega  = H_0 + V_\omega$ on $L^2 ( \R^d
)$, is constructed from a deterministic, periodic, background operator $H_0
= ( - i \nabla - A_0 )^2 + V_0$. We assume that this operator is
self-adjoint with operator core $C_0^\infty ( \R^d )$, and that $H_0
\geq  - M_0 > - \infty$, for some finite constant $M_0$. We consider an
Anderson-type potential $V_\omega$ constructed from the nonzero single-site
potential $u \geq 0$ as
\begin{equation}\label{anderson1}
V_\omega (x) = \Sum_{j \in \Z^d  } \; \omega_j u ( x - j ) .
\end{equation}
We assume very little on the random variables $\{ \omega_j ~| ~j \in
\Z^d \}$ except that they form a bounded, real-valued process over
$\Z^d$ with probability space $( \P , \Omega)$.  We remark that the
results of this paper also apply to the random operators describing
acoustic and electromagnetic waves in randomly perturbed media, and we
refer the reader to \cite{[CHT],[FK1],[FK2]}.

We need to define local versions of the Hamiltonians and potentials
associated with bounded regions in $\R^d$.  By $\Lambda_l (x)$, we
mean the open cube of side length $l$ centered at $x \in \R^d$.  For
$\Lambda \subset \R^d$, we denote the lattice points in $\Lambda$ by
${\tilde \Lambda } = \Lambda \cap \Z^d$.
For a cube $\Lambda$, we take $H_0^\Lambda$ and
$H_\omega^\Lambda$ to be the restrictions of $H_0$ and $H_\omega$,
respectively, to the cube $\Lambda$, with periodic boundary
conditions on the boundary $\partial \Lambda$ of $\Lambda$.
We denote by $E_0^\Lambda ( \cdot )$ and $E_\Lambda ( \cdot )$
the spectral families for
$H_0^\Lambda$ and $H_\omega^\Lambda$, respectively.  Furthermore, for
$\Lambda \subset \R^d$, let $\chi_\Lambda$ be the characteristic
function for $\Lambda$.  The local potential $V_\Lambda$ is defined by
\beq\label{localpot1}
V_\Lambda (x) = V_\omega ( x) \chi_\Lambda (x) ,
\eeq
and we assume this can be written as
\beq\label{localpot2}
V_\Lambda (x) = \Sum_{j \in {\tilde \Lambda }} \; \omega_j u ( x   - j ).
\eeq
For example, if the support of $u$ is contained in a single unit
cube, formula (\ref{localpot2}) holds.
We refer to the discussion in \cite{[CHK1]}
when  the support of $u$ is compact, but not necessarily contained
inside one cube. In this case, $V_\Lambda$ can be
written as in (\ref{localpot2}) plus a boundary term of order $| \partial
\Lambda|$
and hence it does not contribute to the large $|\Lambda|$ limit.
Hence, we may assume (\ref{localpot2}) without any loss of generality.
We will also use the local potential obtained from (\ref{localpot2})
by
setting all the random variables to one, that is,
\beq\label{localpot3}
{\tilde V}_\Lambda (x) = \Sum_{j \in {\tilde \Lambda }} ~u ( x   - j ).
\eeq

We will always make the following four assumptions:
\vspace{.1in}
\begin{description}
\item[(H1).] The background operator $H_0 = (- i \nabla - A_0)^2 + V_0$ is
a lower semi-bounded,
$\Z^d$-perio\-dic \Schr\ operator with a real-valued, $\Z^d$-periodic,
potential
$V_0$, and a $\Z^d$-periodic vector potential $A_0$. We assume that $V_0$
and
$A_0$ are sufficiently regular so that $H_0$ is essentially
self-adjoint on $C_0^\infty ( \R^d)$.

\item[(H2).] The periodic operator $H_0$  has the unique continuation
property, that is,
for any $E \in \R$ and for any function $\phi \in H^2_{loc} ( \R^d)$, if
$(H_0 - E ) \phi = 0$, and if $\phi$ vanishes on an open set, then $\phi
\equiv 0$.

\item[(H3).] The nonzero, nonnegative, compactly-supported,
  single-site potential $u \in L_0^\infty ( \R^d)$, with $\| u
  \|_\infty \leq 1$, and it is strictly positive on a nonempty open
  set.

\item[(H4).] The nonconstant random coupling constants $\{ \omega_j \;
  | \; j \in \Z^d \}$ take values in $[m_0, M_0]$
   and form a real-valued, bounded process $\Z^d$ with probability space $(
\P ,
  \Omega)$.

\end{description}

\noindent
Note that the condition on $\| u \|_\infty$ in (H3) can always be obtained
by rescaling the random variables.

Our main technical result under hypotheses (H1)--(H4) is an optimal
Wegner estimate expressed in Theorem \ref{wegnermain1}. This upper
bound (\ref{wegner11}) is optimal with respect to the volume
dependence and the dependence on the distribution of the random
variables.  This implies the continuity results for the IDS expressed
in Theorems \ref{Lip-main1} -- \ref{landautheorem1}. In order to describe
the
dependence on the probability measure $\P$, we let $\mu_j$ denote the
conditional probability measure for the random variable $\omega_j$ at site
$j \in \Z^d$,
conditioned on all the random variables $(\omega_k)_{k\not=j}$, that is
\begin{equation}\label{genmeas2}
  \mu_j([E,E+\epsilon])=\pro\{\omega_j\in[E,E+\epsilon]\,|
  \,(\omega_k)_{k\not=j}\}
\end{equation}
The Wegner estimate and continuity results for the IDS are expressed in
terms of the
following quantity:
\begin{equation}
  \label{genmeas1}
  s(\epsilon ) \equiv \sup_{j\in\Z^d} \E \left\{
  \sup_{E \in \R} ~\mu_j([E,E+\epsilon]) \right\} .
\end{equation}

Clearly, if the $(\omega_j)_{j\in\Z^d}$ are independent, $\mu_j$ is
just the probability measure of the random variable $\omega_j$. If, in
addition,
the random variables $\omega_j$ are identically distributed,
then all the $\mu_j$ are the same, which we write as $\mu_0$,
and $(\P , \Omega)$ is the usual product probability space.

Our results on the Wegner estimate and the IDS are of greatest interest if
the function $s(\epsilon)$, defined in~\eqref{genmeas1}, satisfies
$s(\epsilon)\to0$, when $\epsilon\to0^+$. In applications to continuity
of the IDS or Anderson localization, the rate of vanishing of
$s(\epsilon)$, as $\epsilon \rightarrow 0^+$, is essential. If, for
example, in the case of independent and identically distributed
(\textit{iid}) random variables, the measure $\mu_j$ is
concentrated on a discrete set, our results do not
provide this control.

We make two  comments on hypotheses (H1)--(H4).
First, concerning the unique continuation property, it is well known
  that $H_0$ has the UCP if $A_0$ and $V_0$ are sufficiently regular;
  e.g. in dimension $d\geq3$, $V_0\in L^{d/2}_{loc}(\R^d)$, $A_0\in
  L^{d}_{loc}(\R^d)$ and $\nabla A_0\in L^{d/2}_{loc}(\R^d)$ are
  sufficient to ensure that $H_0$ has the UCP (see e.g.~\cite{[Wolff]}
  and references therein). It also follows that the Landau Hamiltonian
  (\ref{landau1}) has the UCP.
Second, the boundedness of the random variables is not essential. The
results can be generalized to a class of unbounded random variables.

We define the IDS $N(E)$ for $H_\omega $ using the counting
function for $H_\omega^\Lambda$.
Let $N_\Lambda (E)$ be the number of eigenvalues of $H_\omega^\Lambda$,
with periodic boundary conditions, less than or equal to $E$. This
function depends on the realization $\omega$. The integrated density
of states (IDS) is defined by
\begin{equation*}
N ( E) = \lim_{| \Lambda | \rightarrow \infty } \; \frac{
N_\Lambda (E) }{ | \Lambda | } ,
\end{equation*}
when this limit exists.
As assumptions (H1)-(H4) do not guarantee the existence of this limit,
we will always assume the following.
\begin{description}
\item[(H5).] The IDS $N(E)$ exists almost surely for the random family
  of operators considered here.
\end{description}
Because $N(E)$ is a monotonic function, we assume that $N(E)$ has been
defined to be right continuous, and it has at most a countable number
of discontinuities.  For example, if the family $H_\omega $ is an
ergodic family of random Schr{\"o}dinger operators, it is known that this
limit exists and is independent of the realization $\omega$ almost
surely (cf.\ \cite{[CL],[Kirsch1],[PF]}).  Furthermore, it is known
that the IDS is independent of the boundary conditions taken on the
finite volumes $\Lambda$, cf.\ \cite{[DIM],[Kirsch1],[Nakamura]}.
Our main new result on the IDS is the following theorem.

\vspace{.1in}

\begin{theorem}\label{Lip-main1}
  Assume that the family of random \Schr\ operators $H_\omega = H_0 +
  V_\omega$ on $L^2 ( \R^d)$, for $d \geq 1$,
  satisfies hypotheses (H1)-(H5).
  Then, for any $I\subset\R$ compact, there exists $C_I>0$ such
  that for any $E\in I$ and for any $\epsilon \in (0, 1]$, one has
  \begin{equation*}
    0\leq N(E+\epsilon)-N(E)\leq C_I\,s(\epsilon),
  \end{equation*}
  where $s(\epsilon)$ is defined in~\eqref{genmeas1}.
\end{theorem}

As pointed out above, in order to apply this result to
Anderson localization or to the continuity of the IDS, we need
to impose conditions on the probability measure $\P$ so that
the function $\epsilon\mapsto s(\epsilon)$ vanishes as $\epsilon=0^+$. A
case of
particular interest is when the random variables $(\omega_j)_{j \in \Z^d}$
satisfy not only (H4) but are also
\textit{iid} with
a common probability
measure $\mu_0$ that is locally H{\"o}lder continuous of order $0 < \alpha
\leq 1$.
That is, if for any interval $[a,b] \subset ~\mbox{supp} ~\mu_0$, we
have $\mu_0 ([a,b]) \leq C_0 |b-a|^\alpha$, for some finite, positive
constant $C_0 > 0$ (locally bounded).
The function $s(\epsilon)$ in (\ref{genmeas1}) then
satisfies $s(\epsilon) \leq C_{\mu_0} \epsilon^\alpha$.
Theorem \ref{Lip-main1} states that in this case the IDS $N(E)$ for the
random family $H_\omega $ is locally
H{\"o}lder continuous with uniform H{\"o}lder exponent $\alpha$. That is,
for
any bounded, closed interval $I \subset \R$, there is a finite
positive constant $0 \leq C_I < \infty$, so that for any $E , E' \in
I$, the IDS satisfies
\begin{equation*}
| N(E') - N(E) | \leq C_I |E' - E |^\alpha.
\end{equation*}
If $\alpha = 1$, then the IDS is locally Lipschitz continuous on $\R$.
This condition on the probability measure $\mu_0$
is stronger than just the absolute continuity of the
probability measure as it implies that it admits a nonnegative,
bounded, compactly-supported density $h_0$.
Note that, in the \textit{iid} case, the existence of the IDS is well known,
hence, assumption (H5) can be dropped.
We have the following simple, but important, corollary.

\begin{corollary}\label{Cor-main1}
  Suppose the random family satisfies (H1)-(H3) and the random
  variables $(\omega_j)_{j \in \Z^d}$ are \textit{iid} and the common
   probability measure $\mu_0$ is locally Lipschitz
  continuous and compactly supported. Then the IDS $N(E)$
  is locally uniformly
  Lipschitz continuous and the density of states $\rho (E)$
  exists as a
  locally bounded function.
\end{corollary}

We remark that Corollary \ref{Cor-main1} follows from the new analysis in
section \ref{mainproof} and the spectral averaging result of \cite{[CH1]}
that is valid for a compactly-supported,
Lipschitz continuous probability measure
$\mu_0$. In particular, the new spectral averaging result presented in
Theorem \ref{spectralave1} is not needed for
this case.

We next consider the IDS for random Anderson-type
perturbations of Landau Hamiltonians. The unperturbed operator $H_L (B)$
on $L^2 ( \R^2 )$ has the form
\begin{equation}\label{landau1}
H_L(B) = ( - i \nabla - A)^2, ~~\mbox{where} ~~ A(x_1 , x_2 ) = \frac{B}{2}
( - x_2 ,  x_1 ),
\end{equation}
where $B > 0$ is the magnetic field strength.  The spectrum is pure
point and consists of an increasing sequence of degenerate, isolated
eigenvalues $\{ E_j (B) = (2j + 1)B \; | \; j = 0 , 1, \ldots , \}$ of
infinite multiplicity. The unperturbed Hamiltonian $H_L (B)$
satisfies the unique continuation principle as stated in (H2). The IDS for
this model is a piecewise
constant, monotone increasing function (cf.\ the example in
\cite{[Nakamura]}). The perturbed family of operators is
\begin{equation}\label{landau2}
H_\omega  = H_L(B) +  V_\omega ,
\end{equation}
where $V_\omega$ is the Anderson-type random perturbation given in
(\ref{anderson1}).  It is known that $N(E)$ is locally Lipschitz continuous
in the
following sense. Given an $N > 0$, there is a $B_N >0$ so that for $B
>B_N$, the IDS $N(E)$ is Lipschitz continuous on $(0 , 2(N+1)B )
\backslash \{ E_j (B) \; | \; j = 0 , 1, \ldots , N \}$
\cite{[CH2],[Wang1]}.  Under some additional conditions, Wang
\cite{[Wang2]} also proved that $N(E)$ is smooth outside of a given
Landau level for sufficiently large magnetic field strength.  There
has been some discussion as to the behavior of the IDS at the Landau
energies $E_j (B)$.  If the single-site potential $u$ in (\ref{anderson1})
has support including the unit cube $\Lambda_1 (0)$
and satisfies $u | \Lambda_1 (0) > \epsilon \chi_{\Lambda_1 (0)} > 0$, for
some $\epsilon > 0$, then the IDS is locally Lipschitz continuous at
all energies \cite{[CH2]}.
The following theorem improves \cite{[CHK1]} and \cite{[CHKR]}.
Note that the result holds for any nonzero flux.

\begin{theorem}\label{landautheorem1}
  Let $H_\omega$ be the perturbed Landau Hamiltonian
  (\ref{landau1})-(\ref{landau2}) with magnetic field $B \neq 0$.
  Suppose that this family satisfies (H3)--(H5).
  Then, for any $I\subset\R$ compact, there exists $C_I>0$
  such that for any $E\in I$ and for any $\epsilon \in (0, 1]$, one has
  \begin{equation*}
    0\leq N(E+\epsilon)-N(E)\leq C_I\,s(\epsilon),
  \end{equation*}
where $s(\epsilon)$ is defined in~\eqref{genmeas1}.
\end{theorem}

Of course the remarks following Theorem \ref{Lip-main1}, in particular
Corollary \ref{Cor-main1}, hold for the randomly perturbed Landau
Hamiltonian.

Both main results, Theorems \ref{Lip-main1} and \ref{landautheorem1}, are
proved by establishing a Wegner estimate for the local Hamiltonians
$H_\Lambda$ and
using the identity
\beq\label{measure1}
|N(E + \epsilon) - N(E)| \leq \liminf_{|\Lambda| \rightarrow \infty} \E
\left\{
\frac{1}{|\Lambda|}
Tr E_\Lambda ( [E, E + \epsilon] ) \right\},
\eeq
for $\epsilon$ small enough. We prove a new Wegner estimate in this paper
that holds for general probability measures.
The Wegner estimate is also essential in many proofs of Anderson
localization using the method of multiscale analysis.

\begin{theorem}\label{wegnermain1}
  Assume that the family of random Schr{\"o}dinger operators
  $H_\omega = H_0 +  V_\omega$ on $L^2 ( \R^d)$
  satisfies hypotheses (H1)-(H4).  Then, there exists a locally
  uniform constant $C_W > 0$ such that for any $E_0 \in \R$, and
  $\epsilon \in (0, 1]$, the local Hamiltonians $H_\Lambda$
  satisfy the following Wegner estimate
  \bea\label{wegner11}
  \P \{ \mbox{dist} ( \sigma (H_\Lambda), E_0 ) <
  \epsilon \} & \leq & \E \{ Tr E_\Lambda ( [E_0 - \epsilon, E_0 +
  \epsilon ] ) \}
  \nonumber \\
  &\leq & C_W s( 2 \epsilon) | \Lambda| ,
  \eea
  where $s(\epsilon)$ is defined in~\eqref{genmeas1}.
A similar estimate holds for randomly perturbed Landau Hamiltonians.
\end{theorem}

As an application of our results to a situation involving correlated random
variables, we
consider the family of nonsign definite single-site potentials introduced by
Veseli\'c \cite{[ves1]}.
Let $\Gamma \subset \Z^d$ be a finite set of vectors indexed by $k = 0 ,
\ldots, | \Gamma| < \infty$
(we refer to $k \in \Gamma$).
We consider a family of
bounded, real-valued variables $\alpha_j$, for $j \in \Gamma$.
We assume that $\sum_{j \neq 0} |\alpha_j|
< | \alpha_0|$. This condition guarantees the invertibility of a certain
Toeplitz matrix constructed from the $\alpha_j$. Let $w$ be a single-site
potential
as in (H3) and define a new single-site potential $u$ by
\beq\label{nonsign1}
u(x) = \sum_{j \in \Gamma} \alpha_j w(x-j) .
\eeq
Since the coefficients are not required to have fixed sign, the potential
$u$
is not sign definite. We now construct an Anderson-type random potential
with \textit{iid}
random variables $\omega_j$ as
in (\ref{anderson1}).
Upon substituting the definition of $u$ in (\ref{nonsign1}) into
(\ref{anderson1}),
we can write the potential as
\beq\label{nonsign2}
V_\eta (x) = \sum_{j \in \Z^d} \eta_j u(x-j),
\eeq
where the new family of random variables $\eta_j = \sum_{k \in \Gamma}
\alpha_{j-k} \omega_k$
is no longer independent.
They form a correlated process with finite-range determined by $\Gamma$. It
is easy to compute the conditional probability measure $\mu_j$ for the
random variables $\eta_j$ from
the distribution for the variables $\omega_k$. In particular, if the
single-site
probability distribution $\mu_0$ for $\omega_0$ has a density, then so does
the
conditional probability measure $\mu_j$. Theorem \ref{Lip-main1} applies
to this case and
as a result the IDS is Lipschitz continuous at all energies. Veseli\'c
required
that $u$ have a large support satisfying $u \geq C_0 \chi_{\Lambda_1 (0)}$,
but our results
apply for $u$ as in (H3).

There are very few results on the Wegner estimate for
general processes on $\Z^d$. In the \textit{iid} case,
Stollmann \cite{[Stollmann]} considered a general compactly-supported
probability
measure $\mu_0$ and, using a completely different method, proved a
Wegner estimate of the form (\ref{wegner11}) but with a volume factor
of $|\Lambda|^2$, rather than $|\Lambda|$ as in Theorem
\ref{wegnermain1}.  Stollmann's result can be used to prove Anderson
localization for H{\"o}lder continuous probability measures using the
multiscale analysis but, because of the $| \Lambda|^2$-factor, cannot
be used to study the IDS.  More recently, Hundertmark, Killip,
Nakamura, Stollmann, and Veseli\'c \cite{[HKNSV]} obtained
a bound of the form $s( \epsilon) [ \log (1 / \epsilon) ]^d
| \Lambda|$, improving Stollmann's
bound to the correct volume factor, but under the strong
assumption that $u \geq c_0 \chi_{\Lambda_1(0)}$, the characteristic
function of the unit cube $\Lambda_1(0)$. In Theorem
\ref{wegnermain1}, this covering condition is no longer necessary.
The result in \cite{[HKNSV]} follows from a new exponentially
decreasing bound, in the index $n$, on the $n^{th}$ singular value of
the difference of two semigroups generated by Hamiltonians $H_1$ and
$H_2$ for which the perturbation $H_1 - H_2$ has compact support. This
estimate is used to improve the estimate on the spectral shift
function obtained in \cite{[CHN]}. Using these estimates, the authors
improve the H{\"o}lder continuity of the IDS in the H{\"o}lder continuous
situation studied in \cite{[CHK1]} obtaining $\epsilon [
\log (1 / \epsilon ) ]^d$, in place of $\epsilon^p$, for any $0 < p < 1$.

The contents of this paper are as follows.  We prove Theorem
\ref{wegnermain1}, which implies Theorem \ref{Lip-main1}, in section
\ref{mainproof}, assuming a key spectral averaging result. We
prove this new spectral averaging result for general, compactly-supported
probability measures in section \ref{spave}.
We prove the corresponding
result, Theorem\ref{landautheorem1}, for randomly perturbed
Landau Hamiltonians, in section \ref{idslandau}.
In the first appendix, section \ref{traceclass},
we prove some necessary trace estimates.

Applications of Theorem \ref{wegnermain1} to pointwise
bounds on the expectation of the spectral shift function
are presented in \cite{[CHK3]}.

\vspace{.1in}
\noindent
{\bf Acknowledgments.} We thank M.\ Krishna in pointing out an error in an
earlier version of this paper. The proof of Theorem \ref{spectralave2}
using maximally dissipative operators is due to J.\ H.\ Schenker whom we 
thank
for numerous discussions.


\renewcommand{\thechapter}{\arabic{chapter}}
\renewcommand{\thesection}{\thechapter}

\setcounter{chapter}{2} \setcounter{equation}{0}

\section{Proof of Theorem \ref{Lip-main1}}\label{mainproof}

We now prove Theorem \ref{Lip-main1} via (\ref{measure1}) by proving a
Wegner estimate (\ref{wegner11}).
We always assume that $u$ is nonzero so that $V_\omega$ is nonzero.
Recall that by the
operators $H_0^\Lambda$ and $H_\omega^\Lambda$, we mean the operators
$H_0$ and $H_\omega$ restricted to the cube $\Lambda$ with periodic
boundary conditions. We will often write $H_\Lambda$ for
$H_\omega^\Lambda$. Their spectral families are denoted by
$E_0^\Lambda ( \cdot )$ and $E_\Lambda ( \cdot )$, respectively. In
\cite{[CHK1]}, we proved
\begin{theorem}
  \label{floquet1}
  Let $V: \R^d\to\R$ be a bounded, $\Gamma$-periodic, nonnegative
  function. Suppose that $V>0$ on some open set. Consider a bounded
  interval $I\subset\R$. Then, there exists a finite constant $C (
  I , V) >0$ such that, for any $\Lambda\subset\R^d$ cube with
  integral edges (i.e. vertices in $\Z^d$), one has,
  \begin{equation*}
    \label{initialest2} {\mathbf E}^\Lambda_0(I)V_\Lambda {\mathbf
E}^\Lambda_0(I)
    \geq C(I, V) {\mathbf E}^\Lambda_0(I)
  \end{equation*}
  where $V_{\Lambda}$ is the restriction of $V$ to $\Lambda$.
\end{theorem}
This clearly yields that there exists a constant $C ({\tilde \Delta},
u ) > 0$, independent of $\Lambda$, so that
\begin{equation}
  \label{uc1}
  E_0^\Lambda ({\tilde \Delta }) {\tilde V}_\Lambda
  E_0^\Lambda ({\tilde \Delta })
  \; \geq C ({\tilde \Delta}, u ) E_0^\Lambda ({\tilde
    \Delta }) .
\end{equation}
For a fixed, but arbitrary, $E_0 \in
\R$, let $E_0 \in \Delta \subset {\tilde \Delta}$ be
two closed, bounded intervals
centered on $E_0$, and
let $d_\Delta \equiv ~\mbox{dist} ~( \Delta , {\tilde
\Delta}^c )$. We will always assume that $d_\Delta > 0$.

Preparatory to the proof of Theorem \ref{Lip-main1}, we note that
hypothesis (H3) implies the following.  There exists a finite
  constant $D_0 \equiv D_0 (u , d) >0$, depending only on the
  single-site potential $u$, and the dimension $d \geq 1$, so that for
  all $\Lambda \subset \R^d$,
  \begin{equation}\label{inequal1}
    0 \;  \leq \; \tilde{V}_\Lambda^2 \; \leq  \; D_0 (u , d) {\tilde
V}_\Lambda
    ,
  \end{equation}
  where ${\tilde V}_\Lambda$ is defined in (\ref{localpot3}).
We will use this in the proof.

\vspace{.1in}

\noindent
{\bf Proof of Theorem \ref{Lip-main1} } \\
\noindent
1.  Recalling that $E_\Lambda ( \Delta)$ is a trace class operator,
we need to estimate
\begin{equation}
  \label{trace1}
  \E \{ Tr E_\Lambda ( \Delta) \} .
\end{equation}
We begin with a decomposition relative to the spectral projectors
$E_0^\Lambda ( \cdot )$ for the operator $H_0^\Lambda$. We write
\begin{equation}
  \label{trace2}
  Tr E_\Lambda ( \Delta) = Tr E_\Lambda ( \Delta ) E_0^\Lambda
  ( {\tilde \Delta } ) + Tr E_\Lambda ( \Delta ) E_0^\Lambda
  ( {\tilde \Delta }^c ),
\end{equation}
where the intervals $\Delta \subset {\tilde \Delta}$ satisfy $| \Delta
| < 1$ and $d_\Delta > 0$.  If $\tilde{\Delta}$, and consequently
$\Delta$, lies in a spectral gap of $H_0$, then only the second term
on the right in (\ref{trace2}) contributes and the result follows from
(\ref{expect2}).  Hence, we only need to consider the case when
$\Delta$ does not lie in a spectral gap of $H_0$.

\noindent
2. The term involving ${\tilde \Delta }^c$ is estimated as
follows. Since $E_\Lambda ( \Delta)$ is trace class, let $\{
\phi_m^\Lambda \}$
be the set of normalized eigenfunctions in its range. We expand the trace
in these eigenfunctions and obtain
\beq\label{trace3}
Tr E_\Lambda ( \Delta ) E_0^\Lambda ( {\tilde
\Delta }^c ) = \sum_m \langle \phi_m^\Lambda, E_0^\Lambda (
\tilde{\Delta}^c) \phi_m^\Lambda \rangle.
\eeq
From the eigenfunction equation $(H_\omega^\Lambda - E_m ) \phi_m^\Lambda
= 0$, we easily obtain
\begin{equation*}
- ( H_0^\Lambda - E_m)^{-1} E_0^\Lambda ( \tilde{\Delta}^c)
V_\Lambda
\phi_m^\Lambda = E_0^\Lambda ({\tilde \Delta }^c ) \phi_m^\Lambda.
\end{equation*}
Substituting this into the right side of (\ref{trace3}), and resumming
to obtain a trace, we find
\beq\label{trace31}
Tr E_\Lambda ( \Delta ) E_0^\Lambda ( {\tilde
\Delta }^c ) =  ~\sum_m \langle \phi_m^\Lambda, \left(
V_\Lambda \frac{ E_0^\Lambda ( \tilde{\Delta}^c) }{ (H_0^\Lambda - E_m )^2}
V_\Lambda \right) \phi_m^\Lambda \rangle.
\eeq
We next want to replace the energy $E_m \in \Delta$ in the resolvent in
(\ref{trace31}) by a fixed number, say $-M$, assuming $H_0^\Lambda > -M
>- \infty $. To do this, we define an operator $K$ by
\beq\label{change1}
K \equiv \left(
\frac{ H_0^\Lambda + M }{ H_0^\Lambda - E_m} \right)^2 E_0^\Lambda (
\tilde{\Delta}^c) ,
\eeq
and note that $K$ is bounded, independent of $m$, by
\begin{equation*}
  \| K \| \leq K_0 \equiv
  \left[ 1 +  \frac{2(M + \Delta_+)}{d_\Delta} +
    \frac{(M+ \Delta_+)^2}{d_\Delta^2} \right],
\end{equation*}
where $\Delta = [ \Delta_- , \Delta_+]$.
Now, for any $\psi \in L^2 ( \R^d)$,
\bea\label{upperbd1}
\left\langle \psi,
\frac{ E_0^\Lambda ( \tilde{\Delta}^c) }{ (H_0^\Lambda - E_m )^2}
\psi \right\rangle  & \leq &
  \left\langle \frac{ E_0^\Lambda ( \tilde{\Delta}^c) }{
(H_0^\Lambda + M  )} \psi , K \frac{ E_0^\Lambda ( \tilde{\Delta}^c) }{
(H_0^\Lambda + M  )} \psi \right\rangle \nonumber \\
& \leq & K_0 \left\langle \psi , \frac{ E_0^\Lambda ( \tilde{\Delta}^c) }{
(H_0^\Lambda + M  )^2} \psi \right\rangle \nonumber \\
& \leq & K_0 \left\langle \psi , \frac{ 1 }{
(H_0^\Lambda + M  )^2} \psi \right\rangle,
\eea
since $E_0^\Lambda ( \tilde{\Delta}^c) \leq 1$.
We use the bound (\ref{upperbd1}) on the right in (\ref{trace31}) and
expand the potential. To facilitate this, let
$\chi \geq 0$ be a function of compact support slightly larger than the
support of $u$, and so that $\chi u = u$.
We set $\chi_j ( x) = \chi (x - j)$, for $j \in
\Z^d$.
Returning to (\ref{trace31}), we obtain the bound
\bea\label{trace32}
Tr E_\Lambda ( \Delta ) E_0^\Lambda ( {\tilde
\Delta }^c ) & \leq &  K_0 ~Tr E_\Lambda ( \Delta )  \left(
V_\Lambda \frac{ 1  }{ (H_0^\Lambda + M )^2} V_\Lambda \right) \nonumber \\
& \leq &  K_0 \sum_{i,j \in \tilde{\Lambda}}
  | \omega_i \omega_j| ~\left| ~Tr \left[ u_j E_\Lambda ( \Delta ) u_i \cdot
\left( \chi_i \frac{1}{ (H_0^\Lambda + M )^2} \chi_j \right) \right]
\right|   \nonumber
\\
& \leq &  K_0 \sum_{i,j \in \tilde{\Lambda}}
  ~ \left| ~Tr \left[ u_j E_\Lambda ( \Delta ) u_i \cdot
\left( \chi_i \frac{1}{ (H_0^\Lambda + M )^2} \chi_j \right) \right]
              \right| .
\nonumber \\
& &
\eea

\noindent
3. We divide the double sum in (\ref{trace32}) into two terms: For
fixed $i \in \tilde{\Lambda}$, one sum is over $j \in \tilde{\Lambda}$
for which $\chi_i \chi_j = 0$, and in the second sum is over the remaining
$j \in \tilde{\Lambda}$ so that $\chi_i \chi_j \neq 0$.  For the first sum,
we note that the operator $K_{ij} \equiv \chi_i (H_0^\Lambda + M )^{-2}
\chi_j$ in (\ref{trace31}) is trace class for $d = 1,2,3$. Furthermore,
we prove in Lemma \ref{traceclass2} that the operator $K_{ij}$ is
trace class in all dimensions when $\chi_i \chi_j = 0$, and the trace norm
$\| K_{ij}\|_1$ decays exponentially in $\|i-j\|$ as
\beq\label{tracebd11}
\| K_{ij} \|_1 = \| \chi_i (H_0^\Lambda + M )^{-2}
\chi_j \|_1 \leq C_0 e^{- c_0 \| i - j \|},
\eeq
for positive constants
$C_0, c_0 > 0$ depending on $M$.  To control the second sum in
(\ref{trace32}), we define, for each $i \in \tilde{\Lambda}$, an index
set $\mathcal{J}_i = \{ j \in \tilde{\Lambda} ~|~ \chi_i \chi_j \neq 0 \}$.
We note that $| \mathcal{J}_i|$ depends only on $u$, and is
independent of $i$ and $\Lambda$.  We define an operator
$\tilde{K}_\Lambda$ by
\beq\label{defn11} \tilde{K}_\Lambda \equiv \sum_{i \in
  \tilde{\Lambda} ; j \in \mathcal{J}_i} \chi_j K_{ij} \chi_i .  \eeq
In Lemma \ref{traceclass2}, we prove that for any $m > 0$, and
$\sigma_j > 0$, for $j=0,1,\ldots,m$,
\bea\label{bdtrace321} \left| \sum_{i \in \tilde{\Lambda} ; j \in
    \mathcal{J}_i} Tr ~u_j E_\Lambda ( \Delta ) u_i \cdot K_{ij}
\right| & \leq & \left( \sum_{j = 1}^m \frac{\sigma_j}{2^j \sigma_1
    \cdots \sigma_{j-1}} \right) ~Tr E_\Lambda ( \Delta)
\nonumber \\
&& + \left( \frac{1}{2^m \sigma_1 \cdots \sigma_m} \right)
Tr ~E_\Lambda ( \Delta) \cdot \tilde{K}_\Lambda^{2^m} , \nonumber \\
\eea
and that if $m + 2 > \log d / \log 2$, the operator
$\tilde{K}_\Lambda^{2^m}$ is trace class and $\|
\tilde{K}_\Lambda^{2^m} \|_1 \leq C(\chi,m,d) | \Lambda|$.  We next
choose the $\sigma_j$ in Lemma \ref{traceclass2} so that the term
involving $Tr E_\Lambda ( \Delta)$ in (\ref{bdtrace321}) can be moved
to the left in (\ref{trace2}). Since the coefficient in
(\ref{trace32}) is $K_0$, we choose $\sigma_1 = K_0^{-1}$, and
successively $\sigma_j = K_0^{-2^{j-1}}$.  Then, the coefficient in
(\ref{bdtrace321}) is $( 1 - 2^{-m}) K_0^{-1}$.

\noindent
4. We now return to estimating the right side of (\ref{trace32}).
We have seen that in the disjoint support case, the operator $K_{ij} \in
\mathcal{I}_1$, and in the nondisjoint support
case, we must work with $\tilde{K}_\Lambda^{2^m} \in \mathcal{I}_1$, for $m$
large enough.
We first show how to control the expectation of the trace
on the far right of (\ref{bdtrace321}). For simplicity, we write $n = 2^m$
and recall the sets $\mathcal{J}_{j_k}$ defined in the proof of
Lemma \ref{traceclass2}. First, we write this trace
as
\beq\label{bdtrace331}
Tr ~E_\Lambda ( \Delta) \cdot \tilde{K}_\Lambda^n =
\sum_{i \in \tilde{\Lambda}; j \in \mathcal{J}_{j_{n-1}}}
Tr ~u_{j_n} E_\Lambda ( \Delta) u_i \cdot \tilde{K}(n)_{i j_n}.
\eeq
As in Lemma \ref{traceclass2},
the operator $\tilde{K(n)}_{i j}$ is trace class.
The canonical representation of $\tilde{K(n)}_{ij}$  (where we write $j$ for
$j_n$) is
\begin{equation*}
\tilde{K}(n)_{ij} = \sum_l \mu_l^{(ij)} | \phi_l^{(ij)} \rangle \langle
\psi_l^{(ij)}|
\end{equation*}
where $(\phi_l^{(ij)})_{l}$, $(\psi_l^{(ij)})_{l}$ are orthonormal
families and $\D\sum_l|\mu_l^{(ij)}|<+\infty$.\\
Inserting this into the trace (\ref{bdtrace331}),
we obtain
\bea\label{trace4}
Tr ~E_\Lambda ( \Delta ) \cdot \tilde{K}_\Lambda^n  & \leq &
\sum_{i \in \tilde{\Lambda} ; j \in \mathcal{J}_{j_{n-1}}}
\sum_l \mu_l^{(ij)} \langle \psi_l^{(ij)}, u_j E_\Lambda ( \Delta ) u_i
\phi_l^{(ij)} \rangle \nonumber \\
& \leq & \sum_{i \in \tilde{\Lambda}; j \in \mathcal{J}_{j_{n-1}}} \sum_l
\mu_l^{(ij)} \left\{ \langle \psi_l^{(ij)}, u_j E_\Lambda ( \Delta ) u_j
\psi_l^{(ij)} \rangle + \right. \nonumber \\
& & \left. \langle \phi_l^{(ij)}, u_i E_\Lambda ( \Delta ) u_i \phi_l^{(ij)}
  \rangle \right\} .
\eea
We will prove in section \ref{spave}
below that the expectation of the matrix
elements in (\ref{trace4}) satisfy the following bound
\beq\label{expect1}
\E \{ \langle \psi_l^{(ij)}, u_j E_\Lambda ( \Delta ) u_j
\psi_l^{(ij)} \rangle \} \leq  8 s( | \Delta|) ,
\eeq
where $s( \epsilon)$ is defined in (\ref{genmeas1}).
%
%
It follows from (\ref{trace31})-(\ref{trace4}) and the bound (\ref{expect1})
that
\bea\label{expect2} \E \{ Tr ~E_\Lambda ( \Delta )
\cdot \tilde{K}_\Lambda^n \} & \leq & \sum_{i \in \tilde{\Lambda}; j \in
  \mathcal{J}_{j_{n-1}}} ~C(\chi)
s( | \Delta | ) ~ \| \tilde{K}(n)_{ij}\|_1 \nonumber \\
&\leq & C(\chi, m) s( | \Delta| ) | \Lambda | .  \eea
We use the same technique for the disjoint support terms for which the
exponential decay in the trace norm (\ref{tracebd11}) controls the
double sum to give one factor of $|\Lambda|$.  Returning to
(\ref{trace32}), we obtain
\begin{equation*}
\E (  Tr E_\Lambda ( \Delta ) E_0^\Lambda ( {\tilde
    \Delta }^c ) )  \leq   K_0 C(u,m) s( | \Delta|) | \Lambda | ,
\end{equation*}
plus a term involving $Tr E_\Lambda ( \Delta)$ with a coefficient less than
one from
(\ref{bdtrace321}) that is moved to the left in (\ref{trace2}).

\noindent
5. As for the first term on the right in (\ref{trace2}), we use the
fundamental
assumption (\ref{uc1}).
As in \cite{[CH1]}, we will use the spectral projector $E_0
(\tilde{\Delta})$ of $H_0^\Lambda$ in order to control the trace. We have
\bea\label{maintrace1}
Tr E_\Lambda ( \Delta ) E_0^\Lambda (
{\tilde \Delta } ) & \leq & \frac{1}{C ({\tilde \Delta} , u)}
\left\{ Tr E_\Lambda ( \Delta ) E_0^\Lambda ( {\tilde \Delta } )
  {\tilde V}_\Lambda E_0^\Lambda ({\tilde \Delta} ) \right\}  \nonumber \\
& \leq & \frac{1}{C ({\tilde \Delta} , u)} \left\{ Tr E_\Lambda (
  \Delta ) {\tilde V}_\Lambda E_0^\Lambda ( {\tilde \Delta } )
\right. \nonumber \\
& & \left. - Tr E_\Lambda ( \Delta ) E_0^\Lambda ( {\tilde \Delta }^c )
  {\tilde V}_\Lambda E_0^\Lambda ( {\tilde \Delta } ) \right\}.
\eea
We estimate the second term on the right in (\ref{maintrace1}).
Using the H{\"o}lder inequality
for trace norms, we have, for any $\kappa_0 > 0$,
\bea\label{trace7}
\lefteqn{ | Tr
  E_\Lambda ( \Delta) E_0^\Lambda ( {\tilde \Delta }^c ) {\tilde
    V}_\Lambda E_0^\Lambda ( {\tilde \Delta } ) | } \nonumber \\
& \leq & \; \| E_\Lambda ( \Delta) E_0^\Lambda ( {\tilde \Delta }^c )
\|_2 \; \| {\tilde V}_\Lambda E_0^\Lambda ( {\tilde
  \Delta } ) E_\Lambda ( \Delta) \|_2 \nonumber \\
& \leq & \frac{1}{2 \kappa_0 } Tr E_0^\Lambda ( {\tilde\Delta }^c )
E_\Lambda
( \Delta) + \frac{ \kappa_0 }{2} Tr E_\Lambda ( \Delta) E_0^\Lambda (
{\tilde
  \Delta } ) {\tilde V}_\Lambda^2 E_0^\Lambda
( {\tilde \Delta } )E_\Lambda ( \Delta) . \nonumber \\
& &
\eea
We next estimate the second term on the right in (\ref{trace7}).
Let $D_0$ be the constant in (\ref{inequal1}) so that
${\tilde V}_\Lambda^2 \; \leq \; D_0 {\tilde V}_\Lambda$.
Using this, we find that for any $\kappa_1 > 0$,
\begin{eqnarray*}
\lefteqn{Tr E_\Lambda ( \Delta) E_0^\Lambda ( {\tilde \Delta } )
{\tilde V}_\Lambda^2 E_0^\Lambda ( {\tilde \Delta } ) E_\Lambda ( \Delta)}
  \nonumber \\
  & \leq & D_0 \| E_\Lambda ( \Delta) E_0^\Lambda ( {\tilde
  \Delta } ) {\tilde V}_\Lambda \|_2 ~\|E_0^\Lambda ( {\tilde \Delta } )
  E_\Lambda ( \Delta) \|_2 \nonumber \\
& \leq & \frac{D_0 \kappa_1}{2} Tr E_\Lambda ( \Delta) E_0^\Lambda ( {\tilde
  \Delta } ) {\tilde V}_\Lambda^2 E_0^\Lambda ( {\tilde \Delta } )
  E_\Lambda ( \Delta) + \frac{D_0}{ 2 \kappa_1} Tr E_\Lambda ( \Delta)
   E_0^\Lambda ( {\tilde \Delta } ) .
\end{eqnarray*}
We choose $\kappa_1 = 1 / D_0 > 0$ so that $(1 - D_0 \kappa_1 / 2 ) = 1/2 $.
Consequently, we obtain
\beq\label{trace9}
Tr E_\Lambda ( \Delta) E_0^\Lambda ( {\tilde
  \Delta } ) {\tilde V}_\Lambda^2 E_0^\Lambda ( {\tilde \Delta } )
   E_\Lambda ( \Delta)
  \; \leq \; D_0^2 Tr E_\Lambda
( \Delta) E_0^\Lambda ( {\tilde \Delta } ) .
\eeq
Inserting this into (\ref{trace7}), we find
\beq\label{bound1}
| Tr E_\Lambda ( \Delta) E_0^\Lambda ( {\tilde \Delta }^c )
{\tilde V}_\Lambda E_0^\Lambda ( {\tilde \Delta } ) |
\; \leq \; \frac{1}{2 \kappa_0}
Tr E_0^\Lambda ( {\tilde\Delta }^c ) E_\Lambda
( \Delta) + \frac{ \kappa_0 D_0^2}{2}Tr E_\Lambda ( \Delta)
   E_0^\Lambda ( {\tilde \Delta } ) .
\eeq
As a consequence of (\ref{bound1}),
we obtain for the first term on the right in
(\ref{trace2}),
\bea\label{secondterm1}
\lefteqn{ \left( 1 - \frac{\kappa_0 D_0^2}{2 C( {\tilde \Delta}, u) }
\right)
Tr E_\Lambda ( \Delta ) E_0^\Lambda ( {\tilde \Delta} ) } \nonumber \\
&\leq & \frac{1}{C( {\tilde \Delta}, u) } | Tr E_\Lambda ( \Delta )
      {\tilde V}_\Lambda E_0^\Lambda ( {\tilde \Delta} )| + \frac{1}{2
\kappa_0
      C( {\tilde \Delta}, u) } ~ Tr E_\Lambda ( \Delta )
      E_0^\Lambda ( \tilde{\Delta}^c). \nonumber
\eea
We choose $\kappa_0 = C( {\tilde \Delta} , u) / D_0^2$ so that we have
\beq\label{bound2}
Tr E_\Lambda ( \Delta) E_0^\Lambda ( \tilde{\Delta}) \leq  \frac{2}{C(
{\tilde \Delta}, u) } | Tr E_\Lambda ( \Delta )
      {\tilde V}_\Lambda E_0^\Lambda ( {\tilde \Delta} )| + \frac{ D_0^2}{
C( {\tilde \Delta}, u)^2}
      ~Tr E_\Lambda ( \Delta ) E_0^\Lambda ( \tilde{\Delta}^c).
\eeq
As for the first term on the right in (\ref{bound2}), we use H{\"o}lder's
inequality and write
\bea\label{bound3} \lefteqn{ | Tr E_\Lambda ( \Delta )
  {\tilde V}_\Lambda E_0^\Lambda ( {\tilde \Delta} )| } \nonumber \\
& \leq & \| E_0^\Lambda ( {\tilde \Delta} ) E_\Lambda ( \Delta ) \|_2
~\| E_\Lambda ( \Delta ){\tilde V}_\Lambda E_0^\Lambda ( {\tilde
  \Delta} ) \|_2
\nonumber \\
&\leq& \frac{1}{2 \sigma} \| E_0^\Lambda ( {\tilde \Delta} ) E_\Lambda
( \Delta ) \|_2^2 + \frac{\sigma}{2} \| E_\Lambda ( \Delta ){\tilde
  V}_\Lambda E_0^\Lambda ( {\tilde \Delta} )
\|_2^2   \nonumber \\
&\leq& \frac{1}{2 \sigma} Tr E_0^\Lambda ( {\tilde \Delta} ) E_\Lambda
( \Delta ) + \frac{\sigma}{2} Tr E_0^\Lambda ( {\tilde \Delta} )
{\tilde V}_\Lambda E_\Lambda ( \Delta ) {\tilde V}_\Lambda E_0^\Lambda
( {\tilde \Delta} ) , \eea
for any constant $\sigma > 0$. In light of the coefficient in
(\ref{bound2}), we choose $\sigma = 2 / C( {\tilde \Delta} , u)$ and
obtain from (\ref{bound2}) and (\ref{bound3}),
\bea\label{bound4} Tr E_\Lambda ( \Delta) E_0^\Lambda (
\tilde{\Delta}) &\leq& \frac{4}{C( {\tilde \Delta} , u)^2} Tr
E_0^\Lambda ( {\tilde \Delta} ) {\tilde V}_\Lambda E_\Lambda ( \Delta
) {\tilde V}_\Lambda E_0^\Lambda (
{\tilde \Delta} ) \nonumber \\
& & + \frac{2 D_0^2}{ C( { \tilde \Delta}, u)^2} ~Tr E_\Lambda (
\Delta ) E_0^\Lambda ( \tilde{\Delta}^c).  \eea
The second term on the right in (\ref{bound4}) is bounded above as in
(\ref{trace4}) and (\ref{expect2}).

\noindent
6. We estimate the first term on the right in the last line of
(\ref{bound4}).  Let $f_\Delta \in C_0^\infty ( \R)$ be a smooth,
compactly-supported, nonnegative function $0 \leq f \leq 1$, with
$f_\Delta \chi_\Delta = \chi_\Delta$, where $\chi_\Delta$ is the
characteristic function on $\Delta$.  Note that we can take $|
\mbox{supp} ~f| \sim 1$ so that the derivatives of $f$ are order one.
By positivity, we have the bound
\bea\label{bound5} \lefteqn{ Tr
  E_0^\Lambda ( {\tilde \Delta} ) {\tilde V}_\Lambda E_\Lambda (
  \Delta ) {\tilde V}_\Lambda E_0^\Lambda (
  {\tilde \Delta} ) } \nonumber \\
&=& Tr E_\Lambda ( \Delta ) {\tilde V}_\Lambda E_0^\Lambda ({\tilde
  \Delta} ) {\tilde V}_\Lambda E_\Lambda ( \Delta ) \nonumber \\
&\leq& Tr E_\Lambda ( \Delta ) {\tilde V}_\Lambda f_\Delta (
H_0^\Lambda ) {\tilde V}_\Lambda E_\Lambda ( \Delta ).  \eea
Recall that $\chi_j$ is a compactly-supported function so that $u_j \chi_j =
u_j$.  Upon expanding the potential ${\tilde V}_\Lambda$, the term on
the right in (\ref{bound5}) is
\beq\label{maintrace3} \sum_{j,k \in
  {\tilde \Lambda}} ~Tr ~u_k E_\Lambda ( \Delta ) u_j \cdot \chi_j f_\Delta
( H_0^\Lambda ) \chi_k.  \eeq
The operator $\chi_j f_\Delta( H_0^\Lambda ) \chi_k$ is a nonrandom,
trace class operator. As with the operator $K_{ij}$ in
(\ref{trace32}), it admits a canonical representation
\beq\label{expansion1} \chi_j f_\Delta (H_0^\Lambda ) \chi_k = \sum_l
~\lambda_l^{(jk)} | \phi_l^{(jk)} \rangle \langle \psi_l^{(jk)} | ,
\eeq
for orthonormal functions $\phi_l^{(jk)}$ and $\psi_l^{(jk)}$.  This
operator also satisfies a decay estimate of the type
\beq\label{kerneldecay1} \| \chi_j f_\Delta ( H_0^\Lambda ) \chi_k
\|_1 \leq C_N(f) ( 1 + \| k - j \|^2 )^{-N}, \eeq
for any $N \in \N$ and a finite positive constant depending on
$\|f^{(j)}\|$ independent of $|\Delta|$.  This can be proved using the
Helffer-Sj{\"o}strand formula, see, for example, \cite{[GK]}.  Expanding
the trace in (\ref{maintrace3}) as in (\ref{trace4}), we can bound
(\ref{maintrace3}) from above by
\bea\label{expansion2} Tr E_0^\Lambda
(\tilde{\Delta} ) \tilde{V}_\Lambda E_\Lambda ( \Delta )
\tilde{V}_\Lambda E_0^\Lambda ( \tilde{\Delta} ) & = & \sum_l
\sum_{j,k \in \tilde{\Lambda}} ~\lambda_m^{(jk)} ~ \langle
\psi_l^{(jk)}, u_j E_\Lambda ( \Delta ) u_k
\phi_l^{(jk)} \rangle  \nonumber \\
& \leq & \sum_l \sum_{j,k \in \tilde{\Lambda}}~\lambda_l^{(jk)} \{
\langle \psi_l^{(jk)}, u_j E_\Lambda ( \Delta ) u_j \psi_l^{(jk)}
\rangle \nonumber \\
& & + \langle \phi_l^{(jk)}, u_k E_\Lambda ( \Delta ) u_k
\phi_l^{(jk)} \rangle \} .  \eea
As in (\ref{expect2}),
the expectation of the
matrix elements of the projector $E_\Lambda ( \Delta )$ of the type
occurring in (\ref{expansion2}) are bounded above as
\beq\label{spectralave22}
\E \{ \langle \xi, u_l E_\Lambda ( \Delta )
u_l \xi \rangle \} \leq 8 s( | \Delta | ) ,
\eeq where
$\| \xi \| = 1$, and
$s( \epsilon)$ is
defined in (\ref{genmeas1}).
Given this bound, and the decay bound (\ref{kerneldecay1}), we obtain
\bea\label{expansion3} \E \{ Tr E_0^\Lambda (\tilde{\Delta} )
\tilde{V}_\Lambda E_\Lambda ( \Delta ) \tilde{V}_\Lambda E_0^\Lambda (
\tilde{\Delta} ) \} & \leq & 2 \left( \sum_{j,k \in \tilde{\Lambda}}
  ~\| \chi_j f_\Delta ( H_0^\Lambda) \chi_k \|_1 \right)
~ C_0 (u ) s(  | \Delta | )  \nonumber \\
&\leq & C_1 (u ) s( | \Delta | ) | \Lambda | .  \eea
This estimate, together with estimate (\ref{expect2}) and inequality
(\ref{bound2}), prove that
\beq\label{main1}
\E \{ Tr E_\Lambda ( \Delta) \} \leq C_2 (u ) s(  |  \Delta | ) |
\Lambda | .
\eeq
this proves the Wegner estimate of Theorem \ref{wegnermain1}.  The
results on the IDS in Theorem \ref{Lip-main1} now follows from this
Wegner estimate, the additional H{\"o}lder continuity hypothesis, and the
fact that \beq\label{holder1} s( | \Delta| ) \leq C_3 | \Delta
|^\alpha, \eeq for some locally uniform constant $C_3 > 0$. $\Box$


\renewcommand{\thechapter}{\arabic{chapter}}
\renewcommand{\thesection}{\thechapter}

\setcounter{chapter}{3} \setcounter{equation}{0}

\section{Spectral Averaging for General Probability
Measures}\label{spave}

We now turn to the proof of (\ref{expect1}) and (\ref{spectralave22})
for general probability measures.
As noted after Corollary \ref{Cor-main1} in section \ref{intro},
a local Lipschitz condition on the random variables implies
the existence of a bounded density $h_0 \in L_{loc}^\infty (\R)$ with
compact support.  Hence, this case can be treated by the spectral
averaging method of \cite{[CH1],[CHM],[KS]}. For the general case, we
now present a new one-parameter averaging method.

We consider the one-parameter family of operators $H_\Lambda ( \omega_j ) =
H_{j^\perp}^\Lambda + \omega_j u_j$, where $H_{j^\perp}^\Lambda$
is $H_\Lambda$
with $\omega_j = 0$.
Let $E_0 \in \R$ be fixed and arbitrary. We consider
an interval $\Delta _\epsilon = [E_0 , E_0 + \epsilon ]$,
for some fixed $0 < \epsilon < \infty$. A simple
use of the spectral theorem for a self-adjoint operator $H$
with spectral family $E_H(\cdot)$ shows that
\bea\label{impart01}
\lefteqn{ \int_{\Delta_\epsilon} ~dE ~\langle \phi,
\Im (H-E-i \epsilon)^{-1} \phi
\rangle } \nonumber \\
&=& \langle \phi , \left[
\tan^{-1} \left( \frac{ E_0 + \epsilon - H}{ \epsilon} \right)
- \tan^{-1} \left( \frac{ E_0 - H}{ \epsilon} \right) \right]
   \phi \rangle \nonumber \\
& \geq & (\tan^{-1} 1) \langle \phi, E_H ( \Delta_\epsilon ) \phi \rangle
= (\pi / 4 ) \langle \phi, E_H ( \Delta_\epsilon ) \phi \rangle .
\eea
Applying this to the matrix element in (\ref{spectralave22}), we obtain
\beq\label{rep1}
\langle \phi, u_j E_\Lambda ( \Delta_\epsilon ) u_j \phi \rangle \leq
(\frac{4}{\pi}) \int_{\Delta_\epsilon} ~dE ~\Im ~\langle u_j \phi,
\frac{1}{H_{j^\perp}^\Lambda +
\omega_j u_j - E - i \epsilon } u_j \phi \rangle.
\eeq
Our goal is to evaluate the expectation of the
matrix element in (\ref{rep1}) with respect to the random
variable $\omega_j$. To this end, we prove a new spectral averaging result
that is a discretized version of previous spectral averaging results.
%
%

\begin{theorem}\label{spectralave1}
  Let $A$ and $B$ be two self-adjoint operators on a separable Hilbert
  space $\mathcal{H}$, and suppose that $B$ is bounded and non
  negative.  Then, for any $\phi \in \mathcal{H}$, we have the bound
  \beq\label{me00} \sum_{n \in \Z} ~\sup_{y \in [0,1]} ~\langle B
  \phi, \frac{1}{ (A + (n+y)B)^2 + 1 } B\phi \rangle \leq
  \pi\|B\|(1+\|B\|)\| \phi \|^2.  \eeq
\end{theorem}

The proof of Theorem \ref{spectralave1} uses two technical tools: the
following Lemma \ref{sum00}, the simple proof of which is left to the
reader, and Theorem \ref{spectralave2} that utilizes a basic result
from the theory of maximally dissipative operators, that we briefly
recall below.

For $\kappa \in \R$, and $b > 0$, we
define the function
\beq\label{genmeas3}
\ell( \kappa ; b) \equiv
\sum_{n \in \Z} ~\sup_{y \in [0,1]} ~\frac{b}{ (y + n +
\kappa)^2 + b^2} .
\eeq

\begin{lemma}\label{sum00}
  For $b > 0$, the function $\kappa\mapsto \ell(\kappa; b)$ is
  $\Z$-periodic and satisfies the bound
  \beq\label{ellbd1}
    \sup_{\kappa \in \R} \ell( \kappa; b) \leq
    \pi \left( 1 + \frac{1}{b} \right) .
  \eeq
\end{lemma}

\noindent
Next, we recall a main result in the theory of maximally dissipative
operators (cf.\ \cite{[Naboko],[SNF]}). A closed operator $A$ is
maximally dissipative if $\Im A \geq 0$ and $A$ has no proper dissipative
extension.

\begin{proposition}\label{maxdiss1}
Suppose $A$ is a maximally dissipative operator on a
separable Hilbert space $\mathcal{H}$.
Then, there exists
a Hilbert space $\tilde{\mathcal{H}}$,
containing $\mathcal{H}$ as a subspace, an orthogonal
projection $P: \tilde{\mathcal{H}} \rightarrow \mathcal{H}$,
and a self-adjoint dilation $L$ so that for $z \in \C$ with $\Im z < 0$,
\beq\label{dilation1}
(A - z)^{-1} = P (L - z)^{-1} P^*.
\eeq
\end{proposition}

Note that the signs of the imaginary parts in the denominator of the left
side of (\ref{dilation1}) are the same.
Consequently, the result is valid for an operator $A$ if $-A$ is maximally
dissipative provided $\Im z > 0$.
Also note that under the conditions in Proposition \ref{maxdiss1},
we have $\Im (A - z)^{-1} \leq 0$.

Lemma \ref{sum00} and Proposition \ref{maxdiss1}
allows us to prove the following theorem.

\begin{theorem}\label{spectralave2}
  Let $A$ be a maximally dissipative operator and let $B \geq 0 $ be a
  bounded, nonnegative self-adjoint operator on a separable Hilbert
  space $\mathcal{H}$. Fix $\lambda>0$.  Then, for any $\phi \in
  \mathcal{H}$, we have the bound
\beq\label{me000}
- \sum_{n \in \Z} ~\sup_{y \in [0,1]} ~ \Im \langle B^{1/2} \phi,
\frac{1}{ A + (n+y)B + i\lambda B } B^{1/2} \phi \rangle \leq \pi
\left(1+\frac1\lambda\right)\| \phi \|^2.
\eeq
\end{theorem}

\noindent
{\bf Proof:} Let $\delta > 0$ be a small parameter and set
$B_\delta \equiv B + \delta > \delta$, since $B \geq 0$.
As $B_\delta$ is bounded and invertible, we can write
\beq\label{me1}
\langle B_\delta^{1/2} \phi, \frac{1}{A + (n+y)B_\delta + i\lambda B_\delta
} B_\delta^{1/2} \phi \rangle =
\langle \phi, \frac{1}{ B_\delta^{-1/2}A B_\delta^{-1/2} + (n+y) + i\lambda}
\phi \rangle .
\eeq
Since $B \geq 0$ and bounded,
and $A$ is maximally dissipative, so is $B_\delta^{-1/2}A
B_\delta^{-1/2}$.
Let $P$ and $L$ be the orthogonal projector and self-adjoint dilation
associated with $B_\delta^{-1/2}AB_\delta^{-1/2}$
as in Proposition \ref{maxdiss1}.
Let $\mu_L^\psi$ be the spectral measure for $L$ and the vector $\psi$.
We can write the matrix element in (\ref{me1}) as
\beq\label{me2}
\langle P^* \phi, \frac{1}{ L + (n+y) + i\lambda}
P^* \phi \rangle = \int_{\R} ~d \mu_L^{P^* \phi} (s)
\frac{1}{ s + (n+y) + i\lambda} .
\eeq
Inserting (\ref{me2}) into (\ref{me1}), taking the imaginary part,
summing over $n \in \N$, taking the supremum over $y \in [0,1]$, and
using Fubini's Theorem to intervert summation and integration, we
obtain
\bea\label{me3}
\lefteqn{-\sum_{n \in \Z} ~\sup_{y \in [0,1]} \Im \langle B_\delta^{1/2}
\phi,
\frac{1}{A + (n+y)B_\delta + i\lambda B_\delta
} B_\delta^{1/2} \phi \rangle} \nonumber \\
& \leq  & \int_{\R} ~d \mu_L^{P^* \phi} (s)
~\left(
\sup_{\kappa \in \R} \sum_{n \in \Z} ~\sup_{y \in [0,1]} ~\frac{\lambda}{ (y
+ n + \kappa)^2 + \lambda} \right) .
\eea
By~(\ref{ellbd1}), the right side of (\ref{me3}) is bounded above by
$\pi(1+\lambda^{-1})\| \phi \|^2$ and we obtain the bound
(\ref{me000}) with $B_\delta$ in place of $B$. Now, $B_\delta
\rightarrow B$ in norm, and the resolvent $(A + (n+y)B_\delta +
i\lambda B_\delta )^{-1}$ converges to $(A+(n+y)B + i\lambda B)^{-1}$,
uniformly in $y$. It follows that each term of the series in
(\ref{me000}), with $B_\delta$ in place of $B$, converges to the
corresponding term with $\delta = 0$, and the result follows by
Fubini's Theorem.  $\Box$.

\vspace{.1in}

\noindent
{\bf Proof of Theorem \ref{spectralave1}:}
We derive Theorem~\ref{spectralave1} from Theorem~\ref{spectralave2}.
Pick $0<\lambda<\|B\|^{-1}$. We write the matrix element on the left
in (\ref{me00}) as
\begin{equation*}
  \begin{split}
    \langle B \phi, \frac{1}{ (A + (n+y)B)^2 + 1 } B\phi \rangle &= -
    \Im
    \langle B \phi, \frac{1}{ A + (n+y)B +i } B \phi \rangle   \\
    &= - \Im \langle B\phi, \frac{1}{ [A+(1-\lambda B)i] + (n+y)B +
      i\lambda B } B\phi \rangle .
  \end{split}
\end{equation*}
The operator $A+(1-\lambda B)i$ is maximally dissipative as $A$ is
self-adjoint and $1-\lambda B\geq1-\lambda\|B\|>0$ (see e.g.\ Lemma
B.1 in \cite{[AENSS]}).  We apply Theorem~\ref{spectralave2} with
$\phi$ replaced with $B^{1/2}\phi$ and thus obtain
\begin{equation*}
  \begin{split}
\sum_{n \in \Z} ~\sup_{y \in [0,1]} ~\langle B
  \phi, \frac{1}{ (A + (n+y)B)^2 + 1 } B\phi \rangle& \leq
  \pi(1+\lambda^{-1})\| B^{1/2}\phi \|^2\\ & \leq
  \pi\|B\|(1+\lambda^{-1})\| \phi \|^2.      
  \end{split}
\end{equation*}
Letting $\lambda$ tend to $\|B\|^{-1}$, this immediately
yields~\eqref{me00}. This completes the proof of
Theorem~\ref{spectralave1}.  $\Box$.

\vspace{.1in}

\noindent
We can now prove the necessary estimate on the expectation of the
integral in (\ref{rep1}) for general probability measures.

\begin{proposition}
  \label{generalmeasure1}
  Let $\mu_j$ denote the probability measure of the random variable
  $\omega_j$ conditioned on all the random variables
  $(\omega_k)_{k\not=j}$ and let $s( \epsilon)$ be as defined in
  (\ref{genmeas1}). Assume (H4) is satisfied.
  For any $\epsilon > 0$, let $\Delta_\epsilon \subset \R$ be an
  interval with $| \Delta_\epsilon| = \epsilon$.  We have the
  following bound on the expectation of the energy integral appearing
  in (\ref{rep1}):
  \beq\label{ellbd2}
  \E \left\{ \int_{\Delta_\epsilon} ~dE \int_\R ~d \mu_j (
  \omega_j) ~\Im \langle \phi, u_j \left( \frac{1}{H_{j^\perp}^\Lambda
      + \omega_j u_j - E - i \epsilon} \right) u_j \phi \rangle
      \right\} \leq
  2 \pi s( \epsilon) \| \phi \|^2.  \eeq
\end{proposition}

\noindent
{\bf Proof:}
%
The imaginary part of the matrix element in (\ref{ellbd2}) is
\beq\label{impart1}
\langle u_j \phi, \frac{\epsilon}{(H_{j^\perp}^\Lambda
- E + \omega_j u_j)^2 + \epsilon^2}
u_j \phi \rangle
= \frac{1}{\epsilon} \langle u_j \phi,
\frac{1}{\epsilon^{-2}(H_{j^\perp}^\Lambda - E + \omega_j u_j)^2 + 1}
u_j \phi \rangle .
\eeq
To apply Theorem \ref{spectralave1},
we choose $B = u_j$
and define a self-adjoint operator $A \equiv
\epsilon^{-1}(H_{j^\perp}^\Lambda - E )$
so the matrix element in (\ref{impart1}) may be written as
\beq\label{impart2}
\langle B \phi, \frac{1}{(A + \epsilon^{-1}\omega_j B)^2 + 1}
B \phi \rangle .
\eeq
We divide the integration
over $\omega_j$ into a sum over intervals $[n \epsilon, (n+1) \epsilon ]$,
and change variables letting $\omega_j / \epsilon = n + y$,
so that $y \in [0,1]$.
We then obtain
\bea\label{integral1}
\lefteqn{
\E \left\{ \int_{\R} ~d \mu_j(\omega_j) \langle B \phi, \frac{1}{(A +
\epsilon^{-1}\omega_j B)^2 + 1}
B \phi \rangle \right\} } \nonumber \\
&=& \E \left\{ \sum_n \int_{n\epsilon}^{(n+1) \epsilon} ~d\mu_j(\omega_j)
\langle B \phi, \frac{1}{(A + (n+y) B)^2 + 1}
B \phi \rangle \right\} \nonumber \\
& \leq & \E \left\{ \left( \sup_{m \in \Z} \mu_j ( [ m \epsilon, (m+1)
\epsilon]) \right)
~\sum_n ~\sup_{y \in [0,1]}
\langle B \phi, \frac{1}{(A + (n+y) B)^2 + 1}
B \phi \rangle \right\}  \nonumber \\
  & &
\eea
We apply Theorem \ref{spectralave1} to the last line in (\ref{integral1})
and obtain
\bea\label{integral2}
\E \left\{ \int_{\R} ~d \mu_j(\omega_j) \langle B \phi, \frac{1}{(A +
\epsilon^{-1}\omega_j B)^2 + 1}
B \phi \rangle \right\} & \leq & 2 \pi \|\phi\|^2
~\E \{ [\sup_m \mu_j ( [ m \epsilon, (m+1) \epsilon])] \} \nonumber \\
& \leq & 2 \pi \| \phi \|^2 s(\epsilon) ,
\eea
since $\| B \| = \| u_j \| \leq 1$.
This provides a bound for the average over $\omega_j$ of (\ref{impart1}).
Integrating in energy over $\Delta_\epsilon$, and recalling the factor of
$\epsilon^{-1}$ in (\ref{impart1}), we obtain
the estimate (\ref{ellbd2}). $\Box$


We combine (\ref{impart01}) with (\ref{ellbd2}) to obtain
\beq\label{ellbd3}
\E \{ \langle \phi , u_j E_\Lambda ( \Delta) u_j \phi \rangle \} \leq 8
s( \epsilon ) \| \phi \|^2,
\eeq
which is (\ref{expect1}) and (\ref{spectralave22}).


\renewcommand{\thechapter}{\arabic{chapter}}
\renewcommand{\thesection}{\thechapter}

\setcounter{chapter}{4} \setcounter{equation}{0}

\section{The Integrated Density of States for
Random Landau Hamiltonians}\label{idslandau}

The method of proof in section \ref{mainproof} can be adapted to treat
randomly perturbed Landau Hamiltonians.  The unperturbed Landau
Hamiltonian $H_L (B)$ on $L^2 ( \R^2)$ is described in
(\ref{landau1}), and the perturbed operator $H_\omega $ in
(\ref{landau2}).  The random potential $V_\omega$ is Anderson-type as
in (\ref{anderson1}). A quantitative version of the unique
continuation principle for infinite-volume Landau Hamiltonians,
analogous to (\ref{uc1}), was proved in \cite{[CHKR]}. We note that
this result holds independent of the flux.

\begin{theorem}\label{uclandau1}
Let $H_L (B)$ be the Landau Hamiltonian in (\ref{landau1}) and
let $\Pi_n$ be the projector onto
the infinite-dimensional eigenspace for $H_L(B)$
corresponding to the eigenvalue $E_n(B)$.
Let $u \geq 0$, the single-site potential,
be a nonnegative, compactly-supported
function with $u \in L^\infty ( \R^2 )$, and satisfying $u > u_0 > 0$
on some nonempty open set, for some constant $u_0 > 0$.
We define the potential $\tilde{V}$ by
\begin{equation*}
\tilde{V} (x) \equiv \sum_{j \in \Z^2} u(x-j) .
\end{equation*}
Then, there exists a
finite constant $0 < C_n (B, u) < \infty$, so that
\beq
\label{positive1}
\Pi_n \tilde{V} \Pi_n \; \geq \; C_n ( B , u ) \Pi_n.
\eeq
\end{theorem}

This infinite-volume result was used in \cite{[CHKR]} to prove the
local H{\"o}lder continuity of the IDS, and could be used here to improve
the result to local H{\"o}lder continuity with exponent $0 < \alpha \leq 1$.
However, it is easier to pursue a purely local result and also obtain
a Wegner estimate. Motivated by
transport questions for random Landau Hamiltonians (\ref{landau2}),
Germinet, Klein, and Schenker \cite{[GKS]} used the result (\ref{positive1})
to prove a purely local version of the quantitative unique continuation
principle.
This allowed them to prove a Wegner estimate for Landau Hamiltonians at any
energy,
including the Landau levels.
With this result, we show how to use the method of proof in section
\ref{mainproof} to obtain an improved Wegner estimate and, consequently,
an improved continuity estimate on the IDS.

As in \cite{[GKS]}, given a magnetic field strength $B > 0$,
we define a number $K_B \equiv \mbox{min} \{ k \in \N ~|~ k \geq \sqrt{ B /
4 \pi } \}$, and
a length scale $L_B \equiv K_B \sqrt{B / 4 \pi }$.
Corresponding to $L_B$ we define a set of length scales
$\N_B= L_B \N$.
For squares of side length $L_B N$, the flux is an even integer. The local,
unperturbed Landau Hamiltonians
$H_{\Lambda_L}^0 (B)$ are defined on squares $\Lambda_L
(0)$, with $L \in \N_B$,
with periodic boundary conditions consistent with the magnetic translations.
The spectrum
of these local operators is discrete and consists of finite multiplicity
eigenvalues at the
Landau levels $E_n(B)$. We denote by $\Pi_{n,L}$ the finite rank projection
onto the eigenspace
corresponding to the $n^{th}$ Landau level $E_n(B)$. The local random
Hamiltonians associated with squares
$\Lambda_L (0)$ are defined by $H_\Lambda (B) = H_{\Lambda_L}^0 (B) +
V_\Lambda$, where
\begin{equation*}
V_\Lambda (x) = \sum_{j \in \tilde{\Lambda}_{L - \delta_u} (0)} ~\omega_j
u(x-j),
\end{equation*}
and $\mbox{supp} ~u \subset \Lambda_{\delta_u} (0)$. We obtain local
Hamiltonians
for squares $\Lambda_L (x)$ by conjugation with the magnetic translation
group operators
considered as maps from $L^2(\Lambda_L (0)) \rightarrow L^2 ( \Lambda_L
(x))$.
We always consider $B >0$ fixed.

\begin{theorem}\label{gks1}
\cite{[GKS]} There exists a finite, positive constant $C(n,u) > 0$,
independent of $L \in \N_B$ large enough, so that
\beq\label{uc2}
\Pi_{n,L} {\tilde V}_{\Lambda_L} \Pi_{n,L} \geq C (n,u) \Pi_{n,L} .
\eeq
\end{theorem}

\noindent
We now sketch the proof of the following Wegner estimate from which
the main Theorem \ref{landautheorem1} follows.
The local random Hamiltonians $H_{\Lambda} (B)$ are defined above with $L
\in \N_B$ and
periodic boundary conditions determined by the magnetic translations.

\begin{theorem}\label{wegner1}
We assume hypotheses (H3)-(H4),
and let $I \subset \R$ be a bounded interval.
There is a finite constant $C_W \equiv C_{B, u, I} > 0$,
and a length scale $L_{B, I}$,
so that for any subinterval $\Delta \subset I$ small enough, and
for any $L \in \N_B$ with $L > L_{B, I}$, we have
\begin{equation*}
\E \{ Tr ( E_{\Lambda_L} ( \Delta ) ) \} \;
\leq C_W s( | \Delta |) L^2 ,
\end{equation*}
where $s( \epsilon)$ is defined in (\ref{genmeas1}).
\end{theorem}

\noindent
{\bf Sketch of the Proof of Theorem \ref{wegner1}.} \\
\noindent
1. We write $\Lambda$ for $\Lambda_L$, where $L$ is a
permissible length as described above.
Without loss of generality, we assume that $I$, and the subinterval $\Delta
\subset I$
contains only the Landau
level $E_n(B)$ and no other Landau level. Let $E_0 \in \Delta$
be the center of the interval.
We write the decomposition in (\ref{trace2})
using the unperturbed projector $\Pi_{n,L}$,
\beq\label{ltrace1}
Tr E_\Lambda ( \Delta) = Tr ~E_\Lambda ( \Delta) \Pi_{n,L} +
Tr E_\Lambda ( \Delta ) \Pi_{n,L}^\perp .
\eeq
For the complementary term on the right in (\ref{ltrace1}),
we follow the argument
in (\ref{trace3})--(\ref{trace4}). We can take, for example, $M=1$ in
(\ref{change1}). We easily derive the analog of (\ref{trace32}),
\begin{eqnarray}\label{lexapnd1}
  Tr E_\Lambda (\Delta) \Pi_{n,L}^\perp &\leq & K_n Tr E_\Lambda
  (\Delta)
  V_\Lambda \frac{1}{(H_{\Lambda_L}^0 (B) + 1)^2} V_\Lambda E_\Lambda (
  \Delta) \nonumber \\
  &\leq &  K_n \sum_{i,j \in \tilde{\Lambda}}  Tr \left[ u_j
    E_\Lambda ( \Delta ) u_i \cdot
    \left( \chi_i \frac{1}{ (H_{\Lambda_L}^0 (B) + 1 )^2}
     \chi_j \right) \right] , \nonumber \\
& \\
\end{eqnarray}
where the constant $K_n$ depends on the Landau level $n$ and is expressible
in the form of (\ref{defn11}) with $d_\Delta$ there replaced by
\begin{equation*}
  d_n =  \min ( ~\mbox{dist} ~( I, E_{n-1} (B) ) ,
~\mbox{dist} ~( I, E
_{n+1} (B) ) ).
\end{equation*}
The operator $K_{ij} \equiv \chi_i (H_{\Lambda_L}^0 (B) + 1 )^{-2} \chi_j$
is trace class (cf.\ \cite{[CH2]}) and satisfies an exponential decay
estimate analogous to (\ref{tracebd11}). Completing the argument to
(\ref{expect2}), we obtain
\begin{equation*}
\E \{ Tr E_\Lambda ( \Delta) \Pi_{n,L}^\perp \} \leq  K_n C_0
s(|\Delta|) L^2 .
\end{equation*}

\noindent
2. We now estimate the first term on the right in (\ref{trace1})
using the unique continuation principle (\ref{uc2}),
\bea\label{ltrace3}
Tr E_\Lambda (\Delta) \Pi_{n,L} & \leq & \frac{1}{C(n,u)} \left\{ Tr
E_\Lambda
( \Delta ) {\tilde V}_\Lambda \Pi_{n,L}
                                         \right. \nonumber \\
                       & & \left. - Tr E_\Lambda (\Delta) \Pi_{n,L}^\perp
{\tilde
V}_\Lambda \Pi_{n,L} \right\} .
\eea
We estimate the second term on the right in (\ref{ltrace3})
as in (\ref{trace7})--(\ref{trace9}),
and obtain a bound similar to (\ref{bound1}),
\beq\label{lbound1}
| Tr E_\Lambda ( \Delta ) \Pi_{n,L}^\perp
{\tilde V}_\Lambda \Pi_{n,L} | \leq \frac{1}{2 \kappa_0}
Tr \Pi_{n,L}^\perp E_\Lambda ( \Delta) + \frac{\kappa_0 D_0^2 }{2}
Tr E_\Lambda (\Delta) \Pi_{n,L} ,
\eeq
where we used the constant $D_0$ from (\ref{inequal1}).
We now substitute (\ref{lbound1}) into the right of
(\ref{ltrace3}) and obtain the analog of (\ref{secondterm1}),
\bea\label{lsecondterm1}
\left( 1 - \frac{\kappa_0 D_0^2}{2 C(n,u)} \right) Tr E_\Lambda (\Delta)
\Pi_{n,L}
& \leq & \frac{1}{2 \kappa_0 C(n,u)} Tr  E_\Lambda (\Delta) \Pi_{n,L}^\perp
\nonumber \\
&& + \frac{1}{C(n,u)} | Tr E_\Lambda (\Delta) \tilde{V}_\Lambda
\Pi_{n,L} |  . \nonumber \\
& &
\eea
We choose $\kappa_0 = C(n,u) /  D_0^2$, and obtain from (\ref{lsecondterm1})
an estimate for the left side of (\ref{ltrace3}),
\begin{equation*}
Tr E_\Lambda (\Delta) \Pi_{n,L} \leq \frac{2}{C(n,u)} | Tr E_\Lambda
(\Delta)
{\tilde V}_\Lambda \Pi_{n,L} |
+ \frac{ D_0^2}{ C(n,u)^2} Tr E_\Lambda (\Delta) \Pi_{n,L}^\perp .
\end{equation*}
We follow the same method to estimate the first term on the
right in (\ref{lsecondterm1}) and obtain finally
the analog of (\ref{bound4}),
\begin{equation*}
Tr E_\Lambda (\Delta) \Pi_{n,L}  \leq \frac{2 D_0^2}{C(n,u)^2} Tr E_\Lambda
( \Delta) \Pi_{n,L}^\perp + \frac{4}{C(n,u)^2} Tr \Pi_{n,L}{\tilde
V}_\Lambda
E_\Lambda (\Delta ) {\tilde V}_\Lambda \Pi_{n,L} .
\end{equation*}

\noindent
3. We now estimate $Tr \Pi_{n,L}{\tilde V}_\Lambda E_\Lambda (\Delta)
{\tilde V}_\Lambda \Pi_{n,L}$ as in the proof of Theorem
\ref{Lip-main1}. As in (\ref{bound5}), we first write
\begin{eqnarray*}
Tr \Pi_{n,L}{\tilde V}_\Lambda
E_\Lambda (\Delta) {\tilde V}_\Lambda \Pi_{n,L} &=&
Tr E_\Lambda (\Delta) {\tilde V}_\Lambda \Pi_{n,L} {\tilde V}_\Lambda
  E_\Lambda (\Delta)  \nonumber \\
& \leq & Tr E_\Lambda (\Delta) {\tilde V}_\Lambda f_n (H_{\Lambda_L}^0 (B))
    {\tilde V}_\Lambda \Pi_{n,L} ,
\end{eqnarray*}
where $f_n \in C_0^\infty ( \R)$ is equal to one near $E_n(B)$.
We expand the potential and obtain
\begin{equation*}
  Tr \Pi_{n,L}{\tilde V}_\Lambda
  E_\Lambda (\Delta) {\tilde V}_\Lambda \Pi_{n,L} \leq \sum_{i,j \in
    \tilde{\Lambda}}  ~Tr ~u_j E_\Lambda (\Delta) u_i \cdot \chi_i f_n
  (H_{\Lambda_L}^0 (B))  \chi_j .
\end{equation*}
Following a similar analysis as from (\ref{expansion1}) to
(\ref{expansion3}), we obtain
\begin{equation*}
\E \{ Tr E_\Lambda ( \Delta ) \} \leq C_3(n,u) s( |\Delta| ) L^2,
\end{equation*}
according to hypothesis (H4).  $\Box$


\renewcommand{\thechapter}{\arabic{chapter}}
\renewcommand{\thesection}{\thechapter}

\setcounter{chapter}{6} \setcounter{equation}{0}

\section{Appendix: Trace-class Estimates}\label{traceclass}

For the purposes of this appendix, we let $u \in L^\infty_0 (\R^d)$
denote a compactly-supported function and write $u_j (x) =
u ( x - j)$, for $j \in \Z^d$.
We note that the operator $K_{ij} \equiv u_i  (H_0^\Lambda + M
)^{-2} u_j$ (similar to the operator in (\ref{tracebd11})) is
trace class for $d = 1,2,3$.
For higher dimensions, $d > 3$, we proceed as follows. The operator
$u_i (H_0^\Lambda + M)^{-1} \in \mathcal{I}_q$, where $\mathcal{I}_q$
is the $q^{th}$-von Neumann Schatten class, provided $q > d/2$ (cf.\
\cite{[Simon1]}). We state the essential properties in the following lemma.

\begin{lemma}\label{traceclass2}
Let $u \in L^\infty ( \R^d)$ be a compactly-supported function centered
about the
origin and set $u_j (x) = u(x-j)$, for $j \in \tilde{\Lambda}$, so that
$u_j$
is a compactly-supported function centered about
$j \in \tilde{\Lambda}$.
We assume that $H_0^\Lambda + M$ is boundedly invertible for some $M > 0$,
and for all $\Lambda$.
\begin{enumerate}
\item The bounded operator $K_{ij} \equiv u_i  (H_0^\Lambda + M
)^{-2} u_j$ is trace class if $u_i u_j = 0$. In this case,
there are constants $c_0, C_0 > 0$, independent of $\Lambda$, and $i,j$,
so that
\begin{equation*}
\| K_{ij} \|_1 =
\| u_i (H_0^\Lambda + M )^{-2} u_j \|_1 \leq C_0 e^{- c_0 \| i - j \|} .
\end{equation*}
\item The operator $(H_0^\Lambda + M)^{-1} u_j \in \mathcal{I}_q$, for any
$q > d/2$. Let $\mathcal{J}_i \equiv \{ j \in \tilde{\Lambda} ~| ~u_i u_j
\neq 0 \}$, and define
\begin{equation*}
\tilde{K}_\Lambda \equiv \sum_{i \in \tilde{\Lambda} ;j \in
\mathcal{J}_i} u_i K_{ij} u_j.
\end{equation*}
Then, for any $m > 0$, any  $\sigma_j > 0$, with $\sigma_0 =1$,
we can express the partial sum of the
trace in (\ref{trace32}) in the following form:
\begin{eqnarray*}
\left| \sum_{i \in \tilde{\Lambda} ;j \in
\mathcal{J}_i}
Tr ~E_\Lambda ( \Delta ) \cdot u_i K_{ij} u_j \right|
   & \leq & \left( \sum_{j = 1}^m \frac{\sigma_j}{2^j
\sigma_1 \cdots \sigma_{j-1}}  \right) ~Tr E_\Lambda ( \Delta)
\nonumber \\
&& + \left( \frac{1}{2^m \sigma_1 \cdots \sigma_m} \right)
Tr ~E_\Lambda ( \Delta) \cdot \tilde{K}_\Lambda^{2^m} . \nonumber
\end{eqnarray*}
If $m + 2 > \log d / \log 2$,
the operator $\tilde{K}_\Lambda^{2^m}$ is trace class and
$\| \tilde{K}_\Lambda^{2^m} \|_1 \leq C(u,m,d) | \Lambda|$.
\end{enumerate}
\end{lemma}

\noindent
{\bf Proof.}

\noindent
1. {\it Disjoint Support, Off-Diagonal Terms.}
We first consider separately the terms $K_{ij}$ for which
we have disjoint supports: $u_i u_j = 0$.
Let $R_0 \equiv (H_0^\Lambda + M)^{-1}$ for notational convenience.
Let $\chi$, $\tilde{\chi}$, and $\tilde{\tilde{\chi}}$
be a smooth, compactly-supported function with values in $[0,1]$,
and such that $\chi u = u$.
We choose $\tilde{\chi}$ so that $\tilde{\chi} \chi = \chi$.
We denote by $W(\chi)$ the first-order localized operator $W(\chi)
\equiv [ \chi , H_0 ]$, and we set $\chi_j (x) = \chi(x-j)$, similarly for
$\tilde{\chi}$. If $u_i u_j = 0$, then we can choose $\chi$ and
$\tilde{\chi}$
so that $\chi_j u_i = 0 = \tilde{\chi}_j u_i$. Finally, we take
$\tilde{\tilde{\chi}}_j$ so that $\tilde{\tilde{\chi}}_j W(\tilde{\chi}_j) =
W( \tilde{\chi}_j)$.
In the disjoint support case, we have
\bea\label{disjoint1}
u_i R_0^2 u_j &=& u_i R_0^2 \chi_j u_j \nonumber \\
&=& u_i R_0^2 W(\chi_j) R_0 u_j + u_i R_0 \chi_j R_0 u_j \nonumber \\
&=& u_i R_0^2 W(\chi_j) R_0 u_j + u_iR_0 W(\chi_j) R_0^2 u_j \nonumber \\
&=& u_i R_0^2 W(\tilde{\chi}_j) R_0 W(\chi_j) R_0 u_j + u_i R_0
     W(\tilde{\chi}_j) R_0^2 W(\chi_j) R_0 u_j \nonumber \\
& & + u_i R_0 W( \tilde{\chi}_j) R_0 W(\chi_j) R_0^2 u_j.
\eea
The operator $(H_0^\Lambda + M)^{-1} u_j \in \mathcal{I}_q$, for any
$q > d/2$.
If we suppose that $q = 3$, for example,
then the H{\"o}lder inequality applied to the first term in
(\ref{disjoint1})
implies that
\bea\label{disjoint2}
\| u_i R_0^2 W(\tilde{\chi}_j) R_0 W(\chi_j) R_0 u_j \|_1
  &\leq&  \| u_i R_0 \|_3 ~ \|R_0 W(\tilde{\chi}_j) R_0 W(\chi_j) \|_3
~ \| R_0 u_j \|_3  \nonumber \\
& \leq &  \| u_i R_0 \|_3 ~ \| R_0 \tilde{\tilde{\chi}}_j \|_3 ~ \| R_0 u_j
\|_3 ~\|W(\tilde{\chi}_j) R_0 W(\chi_j) \|
  \nonumber \\
  &<& \infty ,
\eea
since the operator norm on the last line of (\ref{disjoint2}) is bounded.
It is clear that this extends the result to
$d=4,5$. Iterating this scheme with finitely-many cut-off functions, and
recalling that the operator $W(\chi_j) (H_0^\Lambda + M)^{-1} \in
\mathcal{I}_q$, for any
$q > d$, we see that $u_i R_0^2 u_j$ is
trace class in any dimension provided $u_i u_j = 0$.
The exponential decay in the trace norm can be
proved using the Combes-Thomas method, cf.\ \cite{[BCH]}.

\noindent
2. {\it Nondisjoint Support Terms.} Let $\| A \|_2$ denote the
Hilbert-Schmidt norm of an operator $A$.
For $i \in \tilde{\Lambda}$, we let $\mathcal{J}_i \equiv \{ j \in
\tilde{\Lambda} ~| ~u_i u_j
\neq 0\}$, and define
\begin{equation*}
\tilde{K}_\Lambda \equiv \sum_{i \in \tilde{\Lambda} ;j \in
\mathcal{J}_i} u_i K_{ij} u_j.
\end{equation*}
Note that $| \mathcal{J}_i|$ is finite, independent of $i$, depends only on
$\mbox{supp} ~u$,
and so is independent of $|\Lambda|$.
Then we can express the sum of the
nondisjoint support terms occurring in
(\ref{trace32}) in the following form:
\begin{eqnarray*}
  \left| \sum_{i \in \tilde{\Lambda}; j \in \mathcal{J}_i}
    Tr ~u_j E_\Lambda ( \Delta ) u_i \cdot
    K_{ij}  \right|
  &=& | Tr E_\Lambda ( \Delta ) \tilde{K}_\Lambda | \nonumber \\
  &\leq& \| E_\Lambda ( \Delta) \|_2 \| E_\Lambda ( \Delta )
  \tilde{K}_\Lambda  \|_2 \nonumber \\
  &\leq &  \frac{\sigma_1}{2}  Tr E_\Lambda ( \Delta) + \frac1{2\sigma_1}
  Tr E_\Lambda ( \Delta) \tilde{K}_\Lambda^2 ,
\end{eqnarray*}
for any $\sigma_1 > 0$. We iterate this expression $m$ times and obtain
\begin{eqnarray*}
| Tr  E_\Lambda ( \Delta )
   \tilde{K}_\Lambda  | & \leq & \left( \sum_{j=1}^m \frac{\sigma_j}{2^j
\sigma_1 \cdots \sigma_{j-1}} \right) Tr E_\Lambda ( \Delta) \nonumber \\
& & + \left( \frac{1}{ 2^m \sigma_1 \cdots \sigma_m } \right)
Tr ~E_\Lambda ( \Delta) \cdot \tilde{K}_\Lambda^{2^m}
\end{eqnarray*}
where $\sigma_0 \equiv 1$.
To describe the operator $\tilde{K}_\Lambda^n$,
we define an index set $\mathcal{J}_{j_k} \equiv
\{ m \in \tilde{\Lambda} ~| ~u_m u_{j_k} \neq 0 \}$.
We can then write
\begin{equation*}
\tilde{K}_\Lambda^n = \sum_{i \in \tilde{\Lambda}; j_k \in
\mathcal{J}_{j_{k-1}}, k=1, \ldots, n; j_0 = i}
~u_i^2 R_0^2 u_{j_1}^2 u_{j_2}^2 R_0^2 u_{j_3}^2 u_{j_4}^2 R_0^2 \cdots
u_{j_{n-1}}^2 R_0^2 u_{j_n}^2 .
\end{equation*}
Since $u_i R_0^2 u_j \in \mathcal{I}_q$, for any
$q > d/4$, H{\"o}lder's
inequality implies that $\tilde{K}_\Lambda^n \in \mathcal{I}_1$
if $n > d/4$.
It is clear then for $m+2 > \log d / \log 2$,
the operator $\tilde{K}_\Lambda^{2^m}
\in \mathcal{I}_1$. Finally, we easily estimate the trace norm:
\begin{equation*}
\| \tilde{K}_\Lambda^{2^m} \|_1 \leq C(u,m,d) | \Lambda |,
\end{equation*}
for a constant $0 < C(u,m,d) < \infty$ independent of $| \Lambda|$.
$\Box$





\end{document}